\documentclass[twocolumn,prl,amsmath,amssymb,aps,showpacs,superscriptaddress]{revtex4-1}
\usepackage{graphicx}
\usepackage{dcolumn}
\usepackage{bm}
\usepackage{}[multicol]

\newcommand{\eek}{$(e,e^{\prime}K^{+})$}

\newcommand{\henana}{$^{7}_{\Lambda}$He}

\newcommand*{\TOHOKU}{Graduate School of Science, Tohoku University, Sendai, Miyagi 980-8578, Japan}
\newcommand*{\MAINZ}{Institute for Nuclear Physics, Johannes Gutenberg-University, D-55099 Mainz, Germany}
\newcommand*{\NCarolina}{Department of Physics, North Carolina A$\&$T State University,Greensboro, NC 27411, USA}
\newcommand*{\Hampton}{Department of Physics, Hampton University, Hampton, VA 23668, USA}
\newcommand*{\Zagreb}{Department of Physics $\&$ Department of Applied Physics, University of Zagreb, HR-10000 Zagreb, Croatia}
\newcommand*{\Yervan}{A.I.Alikhanyan National Science Laboratory, Yerevan 0036, Armenia}
\newcommand*{\FIU}{Department of Physics, Florida International University, Miami, FL 27411, USA}
\newcommand*{\Christpher}{Department of Physics, Computer Science $\&$ Engineering, Christopher Newport University, Newport News, VA, USA 23606}
\newcommand*{\JLAB}{Thomas Jefferson National Accelerator Facility (JLab), Newport News, VA 23606, USA}
\newcommand*{\Bari}{Istituto Nazionale di Fisica Nucleare, Sezione di Bari and University of Bari, I-70126 Bari, Italy}
\newcommand*{\Southern}{Department of Physics, Southern University at New Orleans,New Orleans, LA 70126, USA}
\newcommand*{\NCarolinatwo}{Department of Physics, University of North Carolina at Wilmington, Wilmington, NC 28403, USA}
\newcommand*{\Roma}{INFN, Sezione Sanit$\grave{a}$ and Istituto Superiore di Sanit$\grave{a}$, 00161 Rome, Italy}
\newcommand*{\Lanzhou}{Nuclear Physics Institute, Lanzhou University, Gansu 730000, China}
\newcommand*{\Mississippi}{Mississippi State University, Mississippi State, Mississippi 39762, USA}
\newcommand*{\James}{Department of Physics and Astronomy, James Madison University, Harrisonburg, VA 22807, USA}
\newcommand*{\Rico}{Escuela de Ciencias y Tecnologia, Universidad Metropolitana, San Juan, 00928, Puerto Rico}
\newcommand*{\VMI}{Department of Physics $\&$ Astronomy, Virginia Military Institute, Lexington, Virginia 24450, USA}
\newcommand*{\Yamagata}{Department of Physics, Yamagata University, Yamagata, 990-8560, Japan}
\newcommand*{\Xavier}{Department of Physics, Xavier University of Louisiana, New Orleans, LA 70125, USA}
\newcommand*{\Hou}{Department of Physics, University of Houston, Houston, Texas 77204, USA}

\begin{document}
\preprint{APS/He7L}
\title{Spectroscopy of the neutron-rich hypernucleus $^{7}_{\Lambda}$He from electron scattering}

\author{T.~Gogami}
\thanks{Current address: {\it Research Center for Nuclear Physics, Osaka University, Ibaraki, Osaka 567-0047, Japan }}
\affiliation{\TOHOKU}
\author{C.~Chen}
\affiliation{\Hampton}
\author{D.~Kawama}
\affiliation{\TOHOKU}
\author{P.~Achenbach}
\affiliation{\MAINZ}
\author{A.~Ahmidouch}
\affiliation{\NCarolina}
\author{I.~Albayrak}
\affiliation{\Hampton}
\author{D.~Androic}
\affiliation{\Zagreb}
\author{A.~Asaturyan}
\affiliation{\Yervan}
\author{R.~Asaturyan}\thanks{Deceased}
\affiliation{\Yervan}
\author{O.~Ates}
\affiliation{\Hampton}
\author{P.~Baturin}
\affiliation{\FIU}
\author{R.~Badui}
\affiliation{\FIU}
\author{W.~Boeglin}
\affiliation{\FIU}
\author{J.~Bono}
\affiliation{\FIU}
\author{E.~Brash}
\affiliation{\Christpher}
\author{P.~Carter}
\affiliation{\Christpher}

\author{A.~Chiba}
\affiliation{\TOHOKU}
\author{E.~Christy}
\affiliation{\Hampton}
\author{S.~Danagoulian}
\affiliation{\NCarolina}
\author{R.~De~Leo}
\affiliation{\Bari}
\author{D.~Doi}
\affiliation{\TOHOKU}
\author{M.~Elaasar}
\affiliation{\Southern}
\author{R.~Ent}
\affiliation{\JLAB}
\author{Y.~Fujii}
\affiliation{\TOHOKU}
\author{M.~Fujita}
\affiliation{\TOHOKU}
\author{M.~Furic}
\affiliation{\Zagreb}
\author{M.~Gabrielyan}
\affiliation{\FIU}
\author{L.~Gan}
\affiliation{\NCarolinatwo}
\author{F.~Garibaldi}
\affiliation{\Roma}
\author{D.~Gaskell}
\affiliation{\JLAB}
\author{A.~Gasparian}
\affiliation{\NCarolina}
\author{Y.~Han}
\affiliation{\Hampton}
\author{O.~Hashimoto}\thanks{Deceased}
\affiliation{\TOHOKU}
\author{T.~Horn}
\affiliation{\JLAB}
\author{B.~Hu}
\affiliation{\Lanzhou}
\author{Ed.V.~Hungerford}
\affiliation{\Hou}
\author{M.~Jones}
\affiliation{\JLAB}
\author{H.~Kanda}
\affiliation{\TOHOKU}
\author{M.~Kaneta}
\affiliation{\TOHOKU}
\author{S.~Kato}
\affiliation{\Yamagata}
\author{M.~Kawai}
\affiliation{\TOHOKU}

\author{H.~Khanal}
\affiliation{\FIU}
\author{M.~Kohl}
\affiliation{\Hampton}
\author{A.~Liyanage}
\affiliation{\Hampton}
\author{W.~Luo}
\affiliation{\Lanzhou}
\author{K.~Maeda}
\affiliation{\TOHOKU}
\author{A.~Margaryan}
\affiliation{\Yervan}
\author{P.~Markowitz}
\affiliation{\FIU}
\author{T.~Maruta}
\affiliation{\TOHOKU}
\author{A.~Matsumura}
\affiliation{\TOHOKU}
\author{V.~Maxwell}
\affiliation{\FIU}
\author{A.~Mkrtchyan}
\affiliation{\Yervan}
\author{H.~Mkrtchyan}
\affiliation{\Yervan}
\author{S.~Nagao}
\affiliation{\TOHOKU}
\author{S.N.~Nakamura}
\affiliation{\TOHOKU}
\author{A.~Narayan}
\affiliation{\Mississippi}
\author{C.~Neville}
\affiliation{\FIU}
\author{G.~Niculescu}
\affiliation{\James}
\author{M.I.~Niculescu}
\affiliation{\James}
\author{A.~Nunez}
\affiliation{\FIU}
\author{Nuruzzaman}
\affiliation{\Mississippi}
\author{Y.~Okayasu}
\affiliation{\TOHOKU}
\author{T.~Petkovic}
\affiliation{\Zagreb}
\author{J.~Pochodzalla}
\affiliation{\MAINZ}
\author{X.~Qiu}
\affiliation{\Lanzhou}
\author{J.~Reinhold}
\affiliation{\FIU}
\author{V.M.~Rodriguez}
\affiliation{\Rico}
\author{C.~Samanta}
\affiliation{\VMI}
\author{B.~Sawatzky}
\affiliation{\JLAB}
\author{T.~Seva}
\affiliation{\Zagreb}
\author{A.~Shichijo}
\affiliation{\TOHOKU}
\author{V.~Tadevosyan}
\affiliation{\Yervan}
\author{L.~Tang}
\affiliation{\Hampton}
\affiliation{\JLAB}
\author{N.~Taniya}
\affiliation{\TOHOKU}
\author{K.~Tsukada}
\affiliation{\TOHOKU}
\author{M.~Veilleux}
\affiliation{\Christpher}
\author{W.~Vulcan}
\affiliation{\JLAB}
\author{F.R.~Wesselmann}
\affiliation{\Xavier}
\author{S.A.~Wood}
\affiliation{\JLAB}
\author{T.~Yamamoto}
\affiliation{\TOHOKU}
\author{L.~Ya}
\affiliation{\Hampton}
\author{Z.~Ye}
\affiliation{\Hampton}
\author{K.~Yokota}
\affiliation{\TOHOKU}
\author{L.~Yuan}
\affiliation{\Hampton}
\author{S.~Zhamkochyan}
\affiliation{\Yervan}
\author{L.~Zhu}
\affiliation{\Hampton}

\collaboration{ HKS (JLab E05-115) Collaboration }

\date{\today}

\begin{abstract}
  The missing mass spectroscopy of the {\henana} hypernucleus
  was performed, using the $^{7}$Li{\eek}{\henana} reaction at 
  the Thomas Jefferson National Accelerator Facility Hall C.
  The $\Lambda$-binding energy of the ground state (1/2$^{+}$)
  was determined with a smaller error
  than that of the previous measurement, being $B_{\Lambda} = 5.55\pm0.10^{\textrm{stat.}}\pm0.11^{\textrm{sys.}}$~MeV.
  The experiment also provided new insight into charge symmetry breaking in $p$-shell hypernuclear systems.
  Finally, a peak at $B_{\Lambda} = 3.65 \pm 0.20^{\textrm{stat.}} \pm 0.11^{\textrm{sys.}}$~MeV
  was observed and  assigned as a mixture of $3/2^{+}$ and $5/2^{+}$ states,
  confirming the ``gluelike'' behavior of $\Lambda$, which makes
  an unstable state in $^{6}$He stable against neutron emission.
\end{abstract}
\maketitle
Nuclear physicists explore the low energy behavior of the
strongly interacting many-body systems, extracting an
effective potential which can be used for nuclear structure and
interaction calculations. Effective potential techniques can also be applied
to hypernuclear systems, as the lifetime of a hyperon in a nucleus
is much greater than the relaxation time associated with strong
interactions. On the other hand, the two-body
potentials for the hyperon-nucleon interaction, $YN$, are not
determined as well as
those for $NN$, due to the experimental difficulties of producing and
detecting hyperons in free scattering experiments.
However, embedding a hyperon within the nuclear
medium (hypernucleus) does allow extraction of
effective potentials from detailed measurements of hypernuclear energy
levels and transitions.

Although many species of $\Lambda$ hypernuclei with masses $A \leq 209$
have been observed~\cite{cite:hashimototamura,cite:feliciello_nagae},
more systematic and precise data are still needed for
further insight into the $\Lambda N$ interaction.
Nowadays, experimental studies of $\Lambda$ hypernuclei use:
(1)~hadron beams at the Japan Proton Accelerator Research Complex
(J-PARC)~\cite{cite:jparc,cite:yamasan},
(2)~heavy ion beams at GSI
~\cite{cite:hyphi,cite:rappold,cite:rappold2},
(3)~heavy-ion colliders at 
the Brookhaven National Laboratory Relativistic Heavy Ion Collider~\cite{cite:star} 
and the CERN Large Hadron Collider~\cite{cite:alice},
and (4)~electron beams at the Mainz Microtron~\cite{cite:patrick, cite:florian} and
the Thomas Jefferson National Accelerator Facility
(JLab)~\cite{cite:miyoshi,cite:lulin,cite:iodice,cite:cusanno,cite:28LAl,cite:cusanno2,cite:7LHe,cite:12LB,cite:9LLi,cite:10LBe}.
The different production mechanisms are complementary and allow us to use their
specific sensitivities to excite particular structures which
highlight nuclear features of interest.

One such feature of interest is charge symmetry breaking (CSB) in
$\Lambda$ hypernuclei.
The difference in ground-state binding energies
in the  $A=3$ nonstrange nuclei ($^{3}$He and $^{3}$H) is
$0.7638 \pm 0.0003$~MeV~\cite{cite:mattauch}.
There remains a binding-energy difference of 0.081 MeV
after accounting for the 0.683~MeV Coulomb correction~\cite{cite:brandenburg}.
In $s$-shell hypernuclei,
a large CSB
$\Delta B_{\Lambda} (^{4}_{\Lambda}{\rm He} - ^{4}_{\Lambda}{\rm H};~0^{+})
= B_{\Lambda}(^{4}_{\Lambda}{\rm He};~0^{+}) - B_{\Lambda}(^{4}_{\Lambda}{\rm H};~0^{+})
= (2.39 \pm 0.03) - (2.04\pm0.04)= +0.35\pm0.06$~MeV
is found by comparing the ground state
binding energies between $^{4}_{\Lambda}$H and $^{4}_{\Lambda}$He~\cite{cite:12LC_2}.
Although the Coulomb effect of the core nuclei $^{3}$He and $^{3}$H 
is already subtracted in the $\Delta B_{\Lambda} (^{4}_{\Lambda}{\rm He} - ^{4}_{\Lambda}{\rm H};~0^{+})$
calculation, the binding-energy difference of the above hypernuclear isospin doublet is due,
in part, to differences in the Coulomb energy caused by contraction of
the nucleus as a result of the additional $\Lambda$ binding.
The Coulomb-energy correction was predicted as
$\Delta B_{C}=0.02$--0.08~MeV~\cite{cite:csbcor1,cite:csbcor2,cite:csbcor3},
and thus $\Delta B_{\Lambda} (^{4}_{\Lambda}{\rm He} -
^{4}_{\Lambda}{\rm H};~0^{+}) + \Delta B_{C} \simeq 0.4$~MeV
is attributed to the $\Lambda N$ CSB for
the $0^{+}$ state in the $A=4$ iso-doublet hypernuclear system.
This difference in the binding energy
is approximately five times larger than
for $A=3$ nonstrange nuclei.
A recent $\gamma$-ray measurement indicates that little binding energy
difference exists between the ($1^{+}$) exited states~\cite{cite:yamasan} 
although it was believed that the $1^{+}$ excited states had as large CSBs 
as the ground states~\cite{cite:bedjidaian1,cite:bedjidaian2,cite:kawachi}.
These residual differences are difficult to explain by Coulomb energy
alone.  
A detailed discussion of hypernuclear CSB~\cite{cite:gal,cite:dani1,cite:dani2}
in addition to other topics of interest have been recently published~\cite{cite:gal_hun_mil}.

CSB in $p$-shell hypernuclear systems
is predicted to be smaller than in $s$-shell systems~\cite{cite:gal}.
Hence, differences in $\Lambda$ binding energies between
$p$-shell mirror hypernuclei are predicted to be less than a few 100~keV ~\cite{cite:gal}.
Previous experiments at JLab Hall C measured $\Lambda$ binding energies of
$^{7}_{\Lambda}$He~\cite{cite:7LHe}, $^{9}_{\Lambda}$Li~\cite{cite:toshi},
$^{10}_{\Lambda}$Be~\cite{cite:10LBe},
$^{12}_{\Lambda}$B~\cite{cite:miyoshi,cite:lulin,cite:12LB},
$^{28}_{\Lambda}$Al~\cite{cite:28LAl}, and $^{52}_{\Lambda}$V~\cite{cite:toshi}
via the {\eek} reaction.
The present paper reports a new result for the $\Lambda$-binding energy
of {\henana} with an improved systematic error and is
compared to its isotopic mirror hypernuclei.
In addition, due to improved statistics, the experiment extracted the first observation of
a peak corresponding to the excited states (3/2$^{+}$, 5/2$^{+}$)
of {\henana}.

CSB in hypernuclear $p$-shell systems can be
studied by comparing the $\Lambda$ binding energies for $A=7$,
isotriplet ($T=1$) $\Lambda$ hypernuclei,
which are the simplest $p$-shell hypernuclear systems,
$^{7}_{\Lambda}$He ($\alpha + n + n + \Lambda$),
$^{7}_{\Lambda}$Li$^{*}$ ($\alpha + p + n + \Lambda$) and
$^{7}_{\Lambda}$Be ($\alpha + p + p + \Lambda$).
The isospin of the ground state of $^{7}_{\Lambda}$Li is $T=0$.
Thus, an excited state of $^{7}_{\Lambda}$Li with $T=1$
should be compared with the iso-triplet partners.
The ground state binding energies of 
$^{7}_{\Lambda}$Li ($T=0$) and $^{7}_{\Lambda}$Be
were measured to be $5.58 \pm 0.03$~MeV and $5.16 \pm 0.08$~MeV, respectively, 
by the emulsion experiments~\cite{cite:juric}.
The binding energy of $^{7}_{\Lambda}$Li$^{*}$ ($T=1$) is 
obtained as $5.26 \pm 0.03$~MeV by using information of 
the energy spacing of $E_{x}(^{7}_{\Lambda}{\rm Li}^{*}; T=1, 1/2^{+})=3.88$~MeV 
measured by the $\gamma$-ray spectroscopy~\cite{cite:tamura} 
and the excitation energy of $E_{x} (^{6}{\rm Li}^{*}; T=1) = 3.56$~MeV~\cite{cite:tilley}.
The ground state (1/2$^{+}$) $\Lambda$-binding
energy of {\henana} using the {\eek} reaction at JLab Hall C (JLab E01-011)
was successfully determined to be
$B_{\Lambda}=5.68 \pm 0.03^{\textrm{stat.}} \pm 0.25^{\textrm{\textrm{sys.}}}$~MeV~\cite{cite:7LHe}.
As a result, the measured energies of $A=7$, $T=1$ hypernuclei
differ from a cluster model prediction which used a phenomenological
$\Lambda N$ CSB potential which was
constructed to reproduce the energies of
$^{4}_{\Lambda}$He and $^{4}_{\Lambda}$H~\cite{cite:hiyama1}.
The error on the $\Lambda$-binding energy
of {\henana} was larger than for other $\Lambda$ hypernuclei
and was dominated by systematic contributions.
Therefore the present experiment (JLab E05-115) focused on the
determination of the $\Lambda$ binding energy of {\henana}
with particular emphasis on reducing the systematic error.

The core nucleus, $^{6}$He ($\alpha + n + n$) in {\henana}
is known as a typical neutron-halo nucleus.
The first excited-state energy of $^{6}$He ($2^{+}$) was measured to be $0.824$~MeV
above the $\alpha + n + n$ breakup threshold, having
a decay width of $\Gamma=0.113$~MeV~\cite{cite:tilley}.
The corresponding states of {\henana} ($3/2^{+}$, $5/2^{+}$) in which
$\Lambda$ resides in the $s$ orbit
are predicted to be stable against neutron-emission
breakup~\cite{cite:hiyama1,cite:richter,cite:hiyama2}
due to the attractive $\Lambda N$ interaction.
In addition, the existence of isomeric states 
in {\henana}~\cite{cite:pniewski1,cite:dalitz,cite:pniewski2}
was speculated from widely scattered binding energy obtained by 
the emulsion experiment,
although it had not been confirmed yet experimentally.
The production cross section for a sum of these states ($3/2^{+}$, $5/2^{+}$)
with the ($\gamma$, $K^{+}$) reaction at the small $K^{+}$ scattering angle
was predicted to be $\approx 60\%$ of that for the ground state
($1/2^{+}$)~\cite{cite:richter}.
Although a small structure was observed in the spectrum which might correspond to
the $3/2^{+}$ and $5/2^{+}$ states, a lack of statistics prevented
confirmation of the observation of these states in
the previous measurement~\cite{cite:7LHe}.
The present experiment acquired five times higher statistics
and can now confirm observation of these states and thus the
``gluelike'' behavior of $\Lambda$. This paper
reports the observation of the ground state ($1/2^{+}$),
and for the first time, the observation of the $3/2^{+}$ and $5/2^{+}$ states.

The {\eek} reaction was used for $\Lambda$ hypernuclear production.
Electroproduction is related to photoproduction
through a virtual photon produced in the ($e,e^{\prime}$)
reaction~\cite{cite:sotona,cite:hunger,cite:xu}.
In the geometry for JLab E05-115,
the virtual photon can be treated as almost real since
the square of the four-momentum transfer $Q^{2}(=-q^{2}>0)$
is quite small [$Q^{2} \simeq$ 0.01~(GeV/$c$)$^{2}$].
The experimental kinematics can be found in Ref.~\cite{cite:12LB}.
We used a continuous wave electron beam
with an energy of $E_{e}=2.344$~GeV, provided by the
Continuous Electron Beam Accelerator Facility at JLab.
The electron beam was transported to the experimental target
which was installed at the entrance of
a charge separation dipole magnet [splitter magnet (SPL)].
The $K^{+}$s ($p^{{\rm center}}_{K^{+}} = 1.200$~GeV/$c$)
and scattered electrons ($p^{{\rm center}}_{e^{\prime}} = 0.844$~GeV/$c$)
were bent in opposite directions by the SPL
and were analyzed with a high-resolution kaon spectrometer
(HKS)~\cite{cite:gogami,cite:fujii}
and a high-resolution electron spectrometer (HES), respectively.
Details for the experimental setup are described
in Refs.~\cite{cite:12LB,cite:10LBe,cite:gogami}.
One important feature of the present experiment
is the excellent resolution of $\Delta p/p \simeq 2 \times 10^{-4}$
(FWHM) for both $K^{+}$ and $e^{\prime}$ at approximately 1~GeV/$c$,
due to the optics of the SPL $+$ HKS $+$ HES system.
Thus, an energy resolution of about 0.5~MeV (FWHM) was obtained
for hypernuclear spectroscopy~\cite{cite:12LB,cite:10LBe}.
The positions and angles of the $K^{+}$s and scattered electrons at
reference planes in the magnetic spectrometers were measured by particle detectors.
This information was converted to momentum vectors at
the target position with backward transfer matrices (BTM)
representing the optical systems for the SPL $+$ HES and SPL $+$ HKS,
respectively,
in order to reconstruct the missing mass ($M_{{\rm HYP}}$).
Once the missing mass was obtained,
the $\Lambda$ binding energy ($B_{\Lambda}$) was calculated as
 $B_{\Lambda} (^{A}_{\Lambda}Z) = M (^{A-1}Z)
  + M_{\Lambda} - M_{{\rm HYP}} (^{A}_{\Lambda}Z)$,
where $Z$ denotes the proton number, and
$M (^{A-1}Z)$ and $M_{\Lambda}$
are the masses of a core nucleus and $\Lambda$.

The energy scale calibration was performed by optimizing
the BTMs of the magnetic spectrometer
systems~\cite{cite:12LB}.
The BTM optimization is also correlated with energy resolution in
the resulting hypernuclear spectra.
For the BTM optimization, events of $\Lambda$ and $\Sigma^{0}$ production
from the 0.45-g/cm$^{2}$ polyethylene target were used along with
events from the production of the $^{12}_{\Lambda}$B ground state
from a 0.0875-g/cm$^{2}$ natural carbon target.
Systematic errors, which originated from the BTM optimization process,
needed to be estimated carefully since the BTM optimization
mainly determines the accuracy of the binding energy ($B_{\Lambda}$)
and excitation energy ($E_{\Lambda}$) of a $\Lambda$ hypernucleus.
In order to estimate the achievable energy accuracy,
a fully modeled Monte Carlo simulation was performed.
As a result,
it was found that $B_{\Lambda}$ and $E_{\Lambda}$
could be determined with accuracies
of $< 0.09$ and $< 0.05$~MeV, respectively, by this optimization method.
Another major contribution to the uncertainty on the binding energy is 
due to energy-loss corrections for the particles in the target. 
This contribution was also estimated by the Monte Carlo simulation 
taking into account the target thickness uncertainty.
Finally, the total systematic errors for $B_{\Lambda}$ and $E_{\Lambda}$ were
estimated as 0.11 and 0.05~MeV, respectively~\cite{cite:12LB,cite:10LBe}.

\begin{figure}[!htbp]
  \begin{center}
    \includegraphics[width=8.6cm]{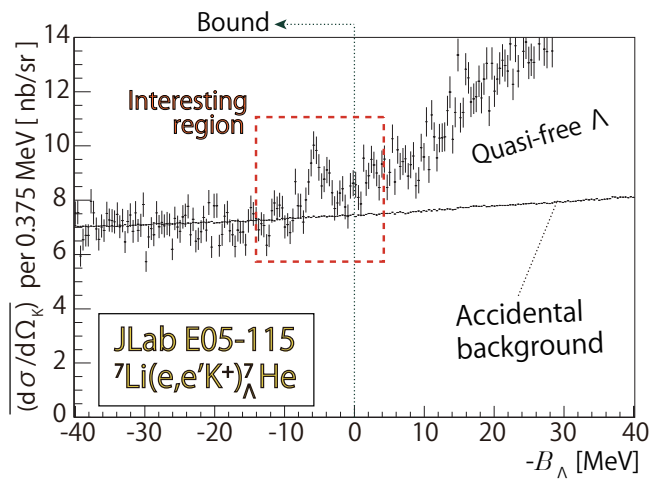}
    \caption{The binding energy ($-B_{\Lambda}$) spectrum of {\henana} with
      an ordinate axis of $\overline{( \frac{d\sigma}{d\Omega_{K}} )}$ defined
      in Ref.~\cite{cite:10LBe}.}
    \label{fig:li7_cs}
  \end{center}
\end{figure}
An enriched $^{7}$Li target (purity of $99\%$) with
a thickness of 0.208~g/cm$^{2}$ was used for the {\henana} production.
The nominal beam current for the production run of {\henana} was
35~$\mu$A, and the total incident charge on the $^{7}$Li target was 4.839~C
($\simeq 3 \times 10^{19}$ electrons).
Figure~\ref{fig:li7_cs} shows the obtained binding energy ($-B_{\Lambda}$) spectrum
with an ordinate axis of $\overline{( \frac{d\sigma}{d\Omega_{K}} )}$
as defined in Ref.~\cite{cite:10LBe}.
For the binding energy calculation, the nuclear masses of
$M (^{7}{\rm Li}) = 5605.54$ and $M (^{6}{\rm He}) = 6533.83$~MeV~\cite{cite:audi}
were used.
Events from quasifree $\Lambda$ production were distributed in the region of $-B_{\Lambda}>0$.
The distribution of the accidental $e^{\prime}K^{+}$-coincidence events in the spectrum was
obtained by the mixed events analysis~\cite{cite:7LHe}.
This method provides an accidental coincidence spectrum with high
statistics thus reducing the statistical uncertainty
caused by background subtraction.

\begin{figure}[!htbp]
  \begin{center}
    \includegraphics[width=8.6cm]{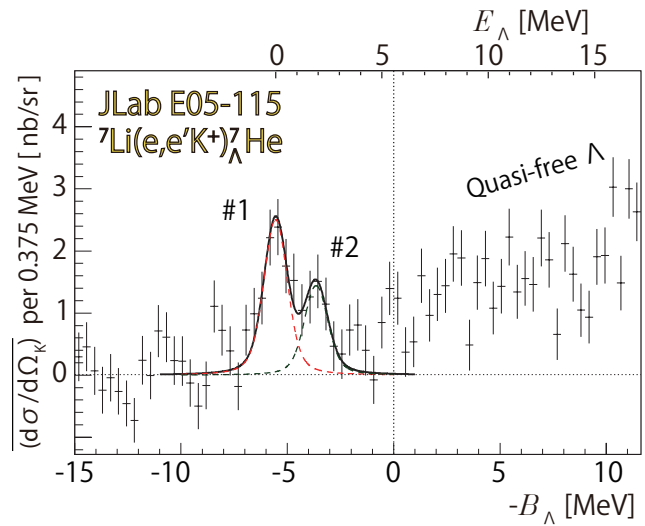}
    \caption{The spectrum of the binding energy ($-B_{\Lambda}$) and the excitation energy
      $\{E_{\Lambda} \equiv -[B_{\Lambda} - B_{\Lambda}(\#1)]\}$
      for the $^{7}$Li{\eek}{\henana} reaction
      with an ordinate axis of $\overline{( \frac{d\sigma}{d\Omega_{K}} )}$
      after the accidental $e^{\prime}K^{+}$-coincidence distribution was subtracted.
      The curve is a fit with two Voigt functions.
    }
    \label{fig:li7csfit}
  \end{center}
\end{figure}
Figure~\ref{fig:li7csfit} shows the spectrum of the $\Lambda$ binding energy ($-B_{\Lambda}$) and
the excitation energy $\{E_{\Lambda} \equiv -[B_{\Lambda}-B_{\Lambda}(\#1)]\}$ of
{\henana} after the accidental $e^{\prime}K^{+}$-coincidence distribution was subtracted.
In order to find peak candidates,
a peak search by tests of statistical significance
defined as $S/\sqrt{S+N}$ was applied.
The statistical significance was calculated for each bin of the histogram,
and the tests for robustness used several settings of bin size
to find peak candidates, taking into account the energy resolution.
As a result, two peak candidates were found with
peak separations of $\geq 3\sigma$ 
as labeled by $\#1$ and $\#2$ in Fig.~\ref{fig:li7csfit}.
The statistical significance for peak $\#1$ is $7.5\sigma$
in a range of $-7.0$ to $-4.0$~MeV,
which is larger than that of the previous measurement ($5.5\sigma$)~\cite{cite:7LHe}.
The two peak candidates were fitted by Voigt functions
(convolution of Gaussian and Lorentzian functions) to obtain
$B_{\Lambda}$ ($E_{\Lambda}$) and the differential cross section for each peak.
The fitting results are given in Table~\ref{tab:LHe7_results}.
\begin{table}[!htbp]
  \begin{center}
    \caption{Fitting results for the $\Lambda$ binding energy,
      excitation energy ($E_{\Lambda}$),
      and $\overline{( \frac{d\sigma}{d\Omega_{K}} )}$
      defined in Ref.~\cite{cite:10LBe} for
      $^{7}$Li{\eek}$^{7}_{\Lambda}$He.
      Errors are statistical and systematic.}
    \label{tab:LHe7_results}
    \begin{tabular}{|c|c|c|c|}
      \hline \hline
      Peak  & Number of events &  $B_{\Lambda}$~(MeV)
      & $\overline{ \Bigl(\frac{d\sigma}{d\Omega_{K}}\Bigr)} $\\
      &      & $[E_{\Lambda}$~(MeV)$]$ & (nb/sr)\\ \hline \hline
      $\#1$ & 413$\pm$38 & 5.55$\pm$0.10$\pm$0.11 & 10.7$\pm$1.0$\pm$1.8\\
        & & (0.0) &\\ \hline
      $\#2$ &  239$\pm$22 & 3.65$\pm$0.20$\pm$0.11 & 6.2$\pm$0.6$\pm$1.1\\
        &  & (1.90$\pm$0.22$\pm$0.05) &\\
      \hline \hline
    \end{tabular}
  \end{center}
\end{table}
The energy resolution was obtained to be 1.3~MeV (FWHM)
which is consistent with the estimation by the Monte Carlo simulation,
although our previously published result for $^{12}_{\Lambda}$B~\cite{cite:12LB} was
better (FWHM $\simeq 0.54$~MeV).
In the Monte Carlo simulation, it was found that our BTMs
have a momentum dependence on the $z$ displacement (beam direction)
from the interaction point. This dependence
significantly contributes to the energy resolution,
adding a kinematical contribution due to the large recoil of
the light hypernuclear system.
The length in the $z$ direction of the $^{7}$Li target (4.0~mm)
was longer than that of the natural $^{12}$C target (0.5~mm)
used for a measurement of $^{12}$C{\eek}$^{12}_{\Lambda}$B~\cite{cite:12LB}.
Thus, the peak width for {\henana} increased
with respect to $^{12}_{\Lambda}$B result as the simulation indicated. 

Peak $\#1$ is considered as the ground state of {\henana}
($^{6}$He$[J_{C};E_{x}]\otimes j^{\Lambda} = [0^{+};{\rm g.s.}] \otimes s^{\Lambda}_{1/2}=$
1/2$^{+}$).
The $\Lambda$ binding energy of the 1/2$^{+}$ state was obtained to be
$5.55 \pm 0.10^{\textrm{stat.}} \pm 0.11^{\textrm{\textrm{sys.}}}$~MeV
which is consistent with the previous result
($5.68 \pm 0.03^{\textrm{stat.}} \pm 0.25^{\textrm{\textrm{sys.}}}$~MeV)~\cite{cite:7LHe}
but with improved uncertainty.
For the previous results,
the statistical error is smaller since 
the energy resolution for $^{7}_{\Lambda}$He spectrum is better 
whereas the systematic error dominates.
In the present result, on the other hand,
statistical and systematic errors are balanced,
reducing the total uncertainty by optimizing
the target thickness and the energy-calibration method.

\begin{figure}[!htbp]
  \begin{center}
    \includegraphics[width=8.6cm]{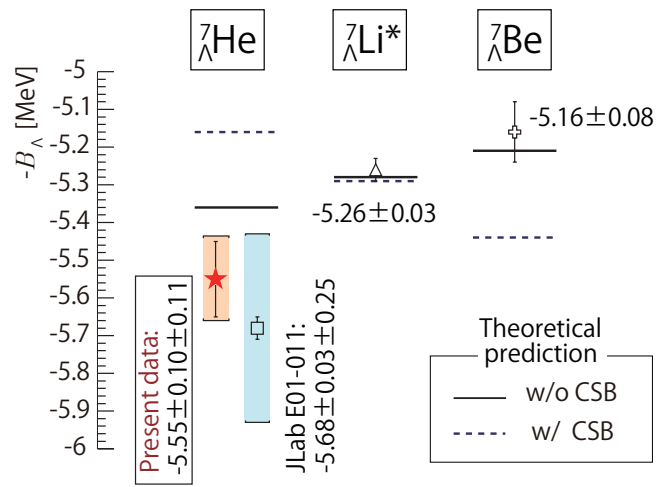}
    \caption{Measured $\Lambda$ binding energies
      of $^{7}_{\Lambda}$He (present result and \cite{cite:7LHe}),
      $^{7}_{\Lambda}$Li$^{*}$~\cite{cite:juric,cite:tamura},
      and $^{7}_{\Lambda}$Be~\cite{cite:juric}
      for the $1/2^{+}$ state with statistical error bars.
      Colored boxes on the experimental results of {\henana}
      indicate systematic errors on $B_{\Lambda}$.
      The solid and dashed lines represent
      theoretical calculations without and with a phenomenological
      even-state $\Lambda N$ CSB potential, which reproduces $\Lambda$ binding
      energies of $^{4}_{\Lambda}$He and $^{4}_{\Lambda}$H,
      by a four-body cluster model~\cite{cite:hiyama1}.
    }
    \label{fig:csb_comparison}
  \end{center}
\end{figure}

Figure~\ref{fig:csb_comparison} shows the measured $\Lambda$ binding energies of
$A=7$, $T=1$ hypernuclei with statistical error bars,
as compared to a theoretical prediction by a four-body cluster model~\cite{cite:hiyama1}.
Colored boxes on the results of {\henana} indicate systematic errors on $B_{\Lambda}$.
In the cluster model prediction~\cite{cite:hiyama1}, a phenomenological
even-state CSB potential was introduced to reproduce the binding energies
of $^{4}_{\Lambda}$He and $^{4}_{\Lambda}$H.
This was applied to the $A=7$, $T=1$ hypernuclear system.
Binding energy predictions without and with the phenomenological CSB potential are
shown by solid and dashed lines, respectively, in Fig.~\ref{fig:csb_comparison}.
The present result seems to favor the energy prediction
without the phenomenological CSB potential.
This is also the case for the other experimental data in the $A=7$, $T=1$ system.
This comparison suggests that
a phenomenological CSB potential needs further consideration.
It is possible to introduce a strong odd-state
CSB potential in addition to one for the even state in order to
make the experiment and theoretical prediction
more consistent~\cite{cite:hiyama1,cite:hiyama3}, although
the validity of a strong odd-state interaction can be questioned~\cite{cite:hiyama4}.
It was suggested that the CSB interaction needs inclusion
of explicit $\Lambda N$-$\Sigma N$ coupling~\cite{cite:gal}.
It seems clear that
further systematic studies, with more precise data particularly for
the $p$-shell hypernuclei are needed.

\begin{figure}[!htbp]
  \begin{center}
    \includegraphics[width=8.6cm]{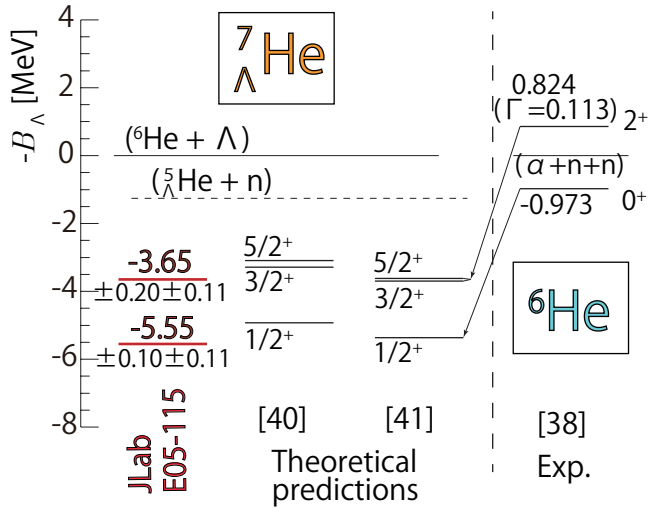}
    \caption{
      Obtained energy levels of {\henana}
      with theoretical predictions~\cite{cite:richter,cite:hiyama2}.
      Reported energy levels for $^{6}$He are also shown~\cite{cite:tilley}.
    }
    \label{fig:enecomp}
  \end{center}
\end{figure}

Peak $\#2$ was obtained at $B_{\Lambda} (\#2) =3.65\pm0.20^{\textrm{stat.}} \pm0.11^{\textrm{sys.}}$~MeV
with a differential cross section of $6.2\pm0.6^{\textrm{stat.}}\pm1.1^{\textrm{sys.}}$~nb/sr.
Figure~\ref{fig:enecomp} shows a $B_{\Lambda}$ comparison
between the obtained results and the theoretical predictions~\cite{cite:richter,cite:hiyama2}
with energy levels of the core nucleus $^{6}$He~\cite{cite:tilley}.
Energy levels of the first excited doublet (3/2$^{+}$, 5/2$^{+}$) are predicted
to be approximately 1.7~MeV above the ground state (1/2$^{+}$)~\cite{cite:richter,cite:hiyama2}.
Moreover, a ratio of the differential cross section of
a sum of $3/2^{+}$ and $5/2^{+}$ states to that of the ground-state ($1/2^{+}$)
is predicted to be approximately 0.6. 
The value of $E_{\Lambda}$ and the ratio of
$\overline{ \Bigl(\frac{d\sigma}{d\Omega_{K}}\Bigr)} $ for peak $\#1$
to peak $\#2$ are
$E_{\Lambda}=1.90\pm0.20^{\textrm{stat.}}\pm0.11^{\textrm{sys.}}$~MeV and $\simeq 0.58$,
respectively.
The results are consistent with the above theoretical predictions, and thus,
peak $\#2$ is interpreted as
$^{6}$He$[J_{C};E_{x}]\otimes j^{\Lambda} = [2^{+};1.8~{\rm MeV}] \otimes s^{\Lambda}_{1/2}=(3/2^{+},5/2^{+})$.

The 3/2$^{+}$ and 5/2$^{+}$ states of {\henana} are more than 2.3~MeV
below the $^{5}_{\Lambda}$He$+n$ breakup energy~\cite{cite:juric,cite:tilley},
which is the lowest neutron emission breakup, as shown in Fig.~\ref{fig:enecomp}.
On the other hand,
the 2$^{+}$ state of $^{6}$He,
which corresponds to the 3/2$^{+}$ and 5/2$^{+}$ states of {\henana},
was reported to be 0.824~MeV above the $\alpha+n+n$ energy~\cite{cite:tilley}
($E=1.797 \pm 0.025$~MeV),
meaning that this state is not stable against neutron emission.
Therefore, the present result of the peak
for the $3/2^{+}$ and $5/2^{+}$ states in {\henana}
confirms the $\Lambda$ gluelike role, making an unbound nucleus bound.

The successful observation of the
first excited doublet in {\henana} opens a door
to study unstable states of light nuclei.
For instance,
energy-level predictions of
the second $2^{+}$ state ($2^{+}_{2}$) in a neutron-halo
nucleus $^{6}$He are largely different depending
on models as shown in Ref.~\cite{cite:mougeot}.
Recently the excitation energy of a $2^{+}_{2}$ state of $^{6}$He was
measured to be $E_{x} = 2.6 \pm 0.3$~MeV
with a width of $\Gamma=1.6 \pm 0.4$~MeV by the
two-neutron transfer reaction $p$($^{8}$He,t)$^{6}$He~\cite{cite:mougeot}.
This measurement would exclude several theoretical models.
However, the $2^{+}_{2}$ energy was derived from a
spectral decomposition by fitting to a spectrum in which a few of
major states are overlapping because of their large decay widths
($\Gamma =$ a few MeV).

On the other hand,
the $3/2_{2}^{+}$ and $5/2_{2}^{+}$ states in {\henana},
which correspond to the $2^{+}_{2}$ state in $^{6}$He,
are predicted to be much narrower~\cite{cite:hiyama2} due to
the additional binding of $\Lambda$.
Therefore, a measurement of the $3/2^{+}_{2}$ and $5/2^{+}_{2}$
states in {\henana}
combined with a realistic cluster calculation
may provide a better understanding of the $2^{+}_{2}$ state in $^{6}$He.
The differential cross section of
the sum of the $3/2^{+}_{2}$ and $5/2^{+}_{2}$ states in {\henana}
is predicted to be approximately $16\%$ of that for the ground state~\cite{cite:hiyama2}.
Consequently, the observation of
the $3/2^{+}_{2}$ and $5/2^{+}_{2}$ states in {\henana}
is promising for future spectroscopy at JLab using the {\eek} reaction.



To summarize, the spectroscopy of $\Lambda$ hypernuclei was performed with
a new magnetic spectrometer system SPL $+$ HKS $+$ HES at
JLab Hall C via the {\eek} reaction.
A spectroscopic measurement of a neutron rich hypernucleus, {\henana},
was performed with an enriched $^{7}$Li target, and
the hypernuclear structure was successfully observed with
an energy resolution of 1.3-MeV FWHM.

The ground state energy ($1/2^{+}$) of {\henana} was determined
to be $B_{\Lambda} = 5.55 \pm 0.10^{\textrm{stat.}} \pm 0.11^{\textrm{sys.}}$~MeV,
which was consistent with the result of the previous measurement
and improved the total error.
The $\Lambda$-binding energy
provides insight into CSB
effects of the $\Lambda N$ interaction
by comparison with the bindings of isotopic mirror
hypernuclei in the $A=7$, $T=1$ system
($^{7}_{\Lambda}$Li$^{*}$, $^{7}_{\Lambda}$Be).
Further systematic investigations with better precision,
particularly for $p$-shell hypernuclei, are necessary
in order to deepen our understanding of $\Lambda N$ CSB.
The {\eek} reaction at JLab
provides a unique method to measure the absolute $\Lambda$-binding energies
of $p$-shell hypernuclei or heavier with less than a few 100-keV accuracy.

The first excited doublet (3/2$^{+}$, 5/2$^{+}$) in
{\henana}, which corresponds to the $2^{+}$ state in $^{6}$He,
was successfully observed for the first time.
A peak for a sum of the 3/2$^{+}$ and 5/2$^{+}$ was
determined to be $B_{\Lambda} = 3.65 \pm 0.20^{\textrm{stat.}} \pm 0.11^{\textrm{sys.}}$~MeV
with the differential cross section of
$\overline{ \Bigl(\frac{d\sigma}{d\Omega_{K}}\Bigr)} = 6.2 \pm 0.6^{\textrm{stat.}} \pm 1.1^{\textrm{sys.}}$~nb/sr.
The peak for the 3/2$^{+}$ and 5/2$^{+}$ states
was found to be approximately 2.3~MeV below the lowest neutron emission energy.
The result shows that the $2^{+}$ state in $^{6}$He,
which is an unstable state for the $\alpha+n+n$ breakup,
becomes stable against neutron-emission breakup
once $\Lambda$ is bound in the nucleus,
owing to the additional binding of $\Lambda$.

We thank the JLab staffs of the physics, accelerator, and
engineering divisions for support of the experiment.
Also, we thank E.~Hiyama, M.~Isaka, D.~J.~Millener, and T.~Motoba
for valuable exchanges related to their theoretical works.
The program at JLab Hall C is supported by JSPS KAKENHI Grants
No.~12002001, No.~15684005, No.~16GS0201, and No.~24$\cdot$4123
(Grant-in-Aid for JSPS fellow),
JSPS Core-to-Core Program No.~21002, and
JSPS Strategic Young Researcher Overseas Visits
Program for Accelerating Brain Circulation No.~R2201.
This work was supported by U.S. Department of Energy Contracts
No.~DE-AC05-84ER40150, No.~DE-AC05-06OR23177, No.~DE-FG02-99ER41065,
No.~DE-FG02-97ER41047, No.~DE-AC02-06CH11357, No.~DE-FG02-00ER41110, and
No.~DE-AC02-98CH10886, and U.S.-NSF Contracts No.~013815 and No.~0758095.


\begin{thebibliography}{00}
\bibitem{cite:hashimototamura} O.~Hashimoto and H.~Tamura, {\it Prog. Part. Nucl. Phys.} {\bf 57}, 564653 (2006).
\bibitem{cite:feliciello_nagae} A.~Feliciello and T.~Nagae, {\it Rep. Prog. Phys. {\bf 78}}, 096301 (2015).
\bibitem{cite:jparc} {\it List of proposed experiments},
[http://j-parc.jp/researcher/Hadron/en/Experiments$\_$e.html]
\bibitem{cite:yamasan}T.O.~Yamamoto {\it et al.} (J-PARC E13 Collaboration), 
{\it Phys. Rev. Lett. {\bf 115}}, 222501 (2015).
\bibitem{cite:hyphi}T.R.~Saito {\it et al.}, {\it Nucl. Phys. A {\bf 881}}, 218--227 (2012).
\bibitem{cite:rappold}C.~Rappold {\it et al.} (HypHI Collaboration), {\it Phys. Rev. C {\bf 88}}, 041001(R) (2013).
\bibitem{cite:rappold2}C.~Rappold, {\it et al.}, {\it Phys. Lett. B {\bf 747}}, 129--134 (2015).
\bibitem{cite:star}Y.~Zhu (for the STAR Collaboration), {\it Nucl. Phys. A {\bf 904--905}}, 551c--554c (2013).
\bibitem{cite:alice}J.~Adam {\it et al.} (ALICE Collaboration), 
{\it Phys. Lett. B {\bf 754}}, 360 (2016).
\bibitem{cite:patrick}A.~Esser, S.~Nagao, F.~Schulz, P.~Achenbach {\it et al.} (A1 Collaboration),
{\it Phys. Rev. Lett. {\bf 114}}, 232501 (2015).
\bibitem{cite:florian}F.~Schulz, P.~Achenbach {\it et al.} (A1 Collaboration), 
{\it Nucl. Phys. A {\bf 954}}, 149 (2016).
\bibitem{cite:miyoshi} T.~Miyoshi {\it et al.} (HNSS Collaboration), {\it Phys. Rev. Lett. {\bf 90}}, 232502 (2003).
\bibitem{cite:lulin} L.~Yuan {\it et al.} (HNSS Collaboration), {\it Phys. Rev. C {\bf 73}}, 044607 (2006).
\bibitem{cite:iodice} M.~Iodice {\it et al.} (Jefferson Lab Hall A Collaboration), {\it Phys. Rev. Lett. {\bf 99}}, 052501 (2007).
\bibitem{cite:cusanno} F.~Cusanno {\it et al.} (Jefferson Lab Hall A Collaboration), {\it Phys. Rev. Lett. {\bf 103}}, 202501 (2009).
\bibitem{cite:28LAl} O.~Hashimoto {\it et al.}, {\it Nucl. Phys. A {\bf 835}}, 121--128 (2010).
\bibitem{cite:cusanno2} F.~Cusanno {\it et al.}, {\it Nucl. Phys. A {\bf 835}}, 129--135 (2010).
\bibitem{cite:7LHe} S.N.~Nakamura, A.~Matsumura. Y.~Okayasu, T.~Seva, V.M.~Rodriguez, P.~Baturin {\it et al.}
(HKS (JLab E01-011) Collaboration), {\it Phys. Rev. Lett. {\bf 110}}, 012502 (2013).
\bibitem{cite:12LB} L.~Tang, C.~Chen, T.~Gogami, D.~Kawama, Y.~Han {\it et al.} (HKS (JLab E05-115 and E01-011) Collaborations), {\it Phys. Rev. C {\bf 90}}, 034320 (2014).
\bibitem{cite:9LLi} G.M.~Urciuoli {\it et al.} (Jefferson Lab Hall A Collaboration), 
{\it Phys. Rev. C {\bf 91}}, 034308 (2015).
\bibitem{cite:10LBe} T.~Gogami, C.~Chen, D.~Kawama {\it et al.} (HKS (JLab E05-115) Collaboration),
{\it Phys. Rev. C {\bf 93}}, 034314 (2016).

\bibitem{cite:mattauch}J.H.E.~Mattauch, W.~Thiele and A.H.~Wapstra, {\it Nucl. Phys. {\bf 67}}, 1 (1965).
\bibitem{cite:brandenburg}R.A.~Brandenburg, S.A.~Coon and P.U.~Sauer, {\it Nucl. Phys. A {\bf 294}}, 3, 305 (1978).
\bibitem{cite:12LC_2} D.H.~Davis, {\it Nucl. Phys. A {\bf 754}}, 3c--13c (2005).
\bibitem{cite:csbcor1}R.H.~Dalitz and F.~Von~Hippel, {\it Phys. Lett. {\bf 10}}, 1 (1964).
\bibitem{cite:csbcor2}J.L.~Friar and B.F.~Gibson, {\it Phys. Rev. C {\bf 18}}, 908 (1978).
\bibitem{cite:csbcor3}A.R.~Bodmer and Q.N.~Usmani {\it Phys. Rev. C {\bf 4}}, 31 (1985).


\bibitem{cite:bedjidaian1}M.~Bedjidian {\it et al.}, {\it Phys. Lett. B {\bf 62}}, 467 (1976).
\bibitem{cite:bedjidaian2}M.~Bedjidian {\it et al.}, {\it Phys. Lett. B {\bf 83}}, 252 (1979).
\bibitem{cite:kawachi}A.~Kawachi, Ph.D. thesis, University of Tokyo, Tokyo, Japan, 1997.
\bibitem{cite:gal} A.~Gal, \textit{Phys. Lett. B {\bf 744}}, 352--357 (2015).
\bibitem{cite:dani1} D.~Gazda and A.~Gal, {\it Phys. Rev. Lett. {\bf 116}}, 122501 (2016).
\bibitem{cite:dani2} D.~Gazda and A.~Gal, {\it Nucl. Phys. A {\bf 954}}, 161 (2016).
\bibitem{cite:gal_hun_mil} A.~Gal, E.V.~Hungerford, and D.J.~Millener, {\it arXiv} 1605.00557,
{\it Rev. Mod. Phys.} (Accepted 25 April 2016).

\bibitem{cite:toshi}T.~Gogami, Ph.D. thesis, Tohoku University, Sendai, Japan, 2014.


\bibitem{cite:juric} M.~Juri$\check{{\rm c}}$ \textit{et al.}, {\it Nucl. Phys. B {\bf 52}}, 1 (1973).
\bibitem{cite:tamura} H.~Tamura {\it et al.}, {\it Phys. Rev. Lett. {\bf 84}}, 5963 (2000).
\bibitem{cite:tilley} D.R.~Tilley, C.M.~Cheves, J.L.~Godwin, G.M.~Hale, H.M.~Hofmann, J.H.~Kelley, C.G.~Sheu,
H.R.~Weller, {\it Nucl. Phys. A {\bf 708}}, 3 (2002).
\bibitem{cite:hiyama1} E.~Hiyama, Y.~Yamamoto, T.~Motoba, M.~Kamimura, \textit{Phys. Rev.} \textbf{C 80}, 054321 (2009).
\bibitem{cite:richter}O.~Richter, M.~Sotona and J.~$\rm{\check{Z}}$ofka, {\it Phys. Rev. C {\bf 43}}, 2753 (1991).
\bibitem{cite:hiyama2}E.~Hiyama, M.~Isaka, M.~Kamimura, T.~Myo, T.~Motoba, {\it Phys. Rev. C {\bf 91}}, 054316 (2015).
\bibitem{cite:pniewski1}J.~Pniewski and M.~Danysz, {\it Phys. Lett. {\bf 1}}, 142 (1962).
\bibitem{cite:dalitz}R.H.~Dalitz and A.~Gal, {\it Nucl. Phys. B {\bf 1}}, 1 (1967).
\bibitem{cite:pniewski2}J.~Pniewski and Z.~Szyma$\acute{{\rm n}}$ski, 
D.H.~Davis, and J.~Sacton, {\it Nucl. Phys. B {\bf 2}}, 317 (1967).

\bibitem{cite:sotona} M.~Sotona and S.~Frullani, \textit{Prog. Theor. Phys. Suppl. {\bf 177}}, 151 (1994).
\bibitem{cite:hunger} Ed V.~Hungerford, \textit{Prog. Theor. Phys. Suppl. {\bf 117}}, 135 (1994).
\bibitem{cite:xu} G.~Xu and Ed V.~Hungerford, \textit{Nucl. Instrum. Methods Phys. Res. A {\bf 501}}, 602--614 (2003).
\bibitem{cite:gogami} T.~Gogami {\it et al.}, {\it Nucl. Instrum. Methods Phys. Res. A {\bf 729}}, 816--824 (2013).
\bibitem{cite:fujii} Y.~Fujii {\it et al.},  {\it Nucl. Instrum. Methods Phys. Res. A {\bf 795}}, 351--363  (2015).

\bibitem{cite:audi} G.~Audi, A.H.~Wapstra and C.~Thibault, \textit{Nucl. Phys. A} \textbf{729}, 337--676 (2003).
\bibitem{cite:hiyama3} E.~Hiyama and Y.~Yamamoto, {\it Prog. Theor. Phys. {\bf 128}}, 1, 105 (2012).
\bibitem{cite:hiyama4} E.~Hiyama, {\it Nucl. Phys. A {\bf 914}}, 130 (2013).
\bibitem{cite:mougeot} X.~Mougeot {\it et al.}, {\it Phys. Lett. B {\bf 718}}, 441 (2012).
\end{thebibliography}
\end{document}